# Stabilization of charge ordering in $La_{1/3}Sr_{2/3}FeO_{3-\delta}$ by magnetic exchange

R. J. McQueeney[1,2], J. Ma[1,2], S. Chang[2], J.-Q. Yan[2], M. Hehlen[3], and F. Trouw[3]

1 - Dept. of Physics and Astronomy, Iowa State University, Ames, IA 50011
2 - Ames Laboratory, Ames, IA 50011
3 - LANSCE, Los Alamos National Laboratory, Los Alamos, NM 87545

## ABSTRACT

The magnetic exchange energies in charge ordered $La_{1/3}Sr_{2/3}FeO_{3-\delta}$ (LSFO) and its parent compound $LaFeO_3$ (LFO) have been determined by inelastic neutron scattering. In LSFO, the measured ratio of ferromagnetic exchange between $Fe^{3+}$ - $Fe^{5+}$ pairs ($J_F$) and antiferromagnetic exchange between $Fe^{3+}$ - $Fe^{3+}$ pairs ($J_{AF}$) fulfills the criterion for charge ordering driven by magnetic interactions ($|J_F/J_{AF}| > 1$). The 30% reduction of $J_{AF}$ as compared to LFO indicates that doped holes are delocalized, and charge ordering occurs without a dominant influence from Coulomb interactions.

The varied, and sometimes extreme, properties of transition metal oxides arise from the common magnetic, vibrational, electronic, and orbital energy scales, leading to complex phase diagrams. One common phase observed in transition metal oxides is the charge ordered ground state. Charge ordered, or nearly charge ordered ground states, are thought to play an important role in colossal magnetoresistive manganites [1] and high-temperature cuprate superconductors (in the form of stripes) [2]. Charge ordering (CO) transitions resulting in a lower electrical conductivity and a change in lattice symmetry are often called Verwey transitions and occur in mixed valent systems such as $Fe_3O_4$ [3], $YBaFe_2O_5$ [4], $La_{1/3}Sr_{2/3}FeO_3$ [5], and $Fe_2OBO_3$ [6]. It is natural to think that CO arises from competition between the Coulomb repulsion energy of localized electrons and kinetic energy cost of forming electronic bands (Wigner crystallization) [7]. This appears to be the main driver for the classic Verwey transition in $Fe_3O_4$ [7,8] and also $Fe_2OBO_3$ [6]. However, the Verwey transition can be strongly influenced by magnetism and magnetic long-range-order. In doped Mott-Hubbard insulators, the doped holes can phase separate and order to form magnetic domain walls that minimize the magnetic exchange energy [9]. In some cases the magnetic energy, rather than the electrostatic energy, may be the dominant factor in the Verwey transition.

    The perovskite compound $La_{1/3}Sr_{2/3}FeO_3$ (LSFO) has been proposed as a system where magnetic interactions drive the Verwey transition [10]. LSFO is derived from the charge-transfer type antiferromagnetic insulator $LaFeO_3$ (LFO, $T_N = 738$ K) [11] by replacement of $La^{3+}$ with $Sr^{2+}$ (increasing the formal Fe valence from 3+ to 3.67+). At temperatures below $T_V = 210$ K (the Verwey transition temperature), charge disproportionation occurs on the Fe sites, $3Fe^{3.67+} \rightarrow 2Fe^{3+} + Fe^{5+}$ [4] leading to

simultaneous charge and spin ordering. The $Fe^{5+}$ valence state is unusual but, due to the small charge transfer gap in LSFO, appreciable hole density on oxygen ions leads to an admixture of additional electronic configurations ($Fe^{5+} \rightarrow Fe^{4+}\underline{L} \rightarrow Fe^{3+}\underline{L}^2$), where $\underline{L}$ represents a hole on the oxygen ligand [12]. It is expected that the oxygen character of the doped holes and the large Fe - O hybridization screens intersite Coulomb interactions significantly and minimizes their influence on CO. Formation of the CO ground state is then driven primarily by the magnetic energy cost of arranging charges to form a metal-centered magnetic domain wall (MCDW). The MCDW consists of an arrangement of $Fe^{5+}$ and $Fe^{3+}$ spin and valence states with ordering vectors $\mathbf{q}_{CO} = (1/3,1/3,1/3)$ [13] and $\mathbf{q}_{AF} = (1/6,1/6,1/6)$ [14], as shown in Fig. 1. According to the Goodenough-Kanamori rules [15], superexchange interactions between $Fe^{3+}$ - $Fe^{3+}$ ions are antiferromagnetic ($J_{AF}$) and $Fe^{5+}$ - $Fe^{3+}$ are ferromagnetic ($J_F$), establishing a consistency between the charge and spin ordered structures (see Fig. 1).

Different domain wall patterns are possible, and each has different magnetic and electrostatic energy cost. When magnetic energies are dominant, the (111) MCDW described above occurs if $|J_F/J_{AF}| > \sim 1$. If $|J_F/J_{AF}| < \sim 1$, magnetic energy will favor a different domain wall structure where planes of 5+ and 3+ valence alternate along the (100) direction [10]. The (111) MCDW is observed experimentally, thus the theory predicts that the ratio $|J_F/J_{AF}|$ must be greater than one when magnetic interactions are dominant. We used inelastic neutron scattering to determine the values of $J_F$ and $J_{AF}$ by measurement of the magnetic excitation spectrum. The measured exchange ratio ($|J_F/J_{AF}| = 1.5$) fulfills the criterion for the formation of a CO ground-state due to magnetic interactions only. In addition, comparison of the measured exchange energies in LSFO to the parent compound LFO are consistent with a scenario where doped oxygen holes are strongly hybridized with iron and delocalized (see Fig. 1). These two results point to a Verwey transition occurring without strong influence from Coulomb interactions.

Inelastic neutron scattering measurements were performed on powders of LSFO and LFO using the Pharos spectrometer at the Lujan Center of Los Alamos National Laboratory. Powders were prepared by conventional solid-state reaction method and subsequently annealed to tune oxygen stoichiometry. Samples weighed approximately 50 grams each and were confirmed to be single-phase by x-ray powder diffraction measurements. Magnetization measurements show the transition to an antiferromagnetic state at $T_V = 210$ K for LSFO. In addition, thermogravimetric analysis and iodometric titration indicate an oxygen stoichiometry parameter $\delta < 0.05$ for LSFO. Powders were packed in flat aluminum cans oriented at $45^o$ or $135^o$ to the incident neutron beam and inelastic neutron scattering spectra were measured over a wide scattering angle range with incident energies of 120, 160, and 300 meV. The time-of-flight data were reduced into energy ($\hbar\omega$) and scattering angle ($2\theta$) histograms and corrections for detector efficiencies, empty can scattering, and instrumental background were performed.

Figure 2(a) shows the full spectrum for LFO at $T = 10$ K as a function of angle and neutron energy loss. In LFO at low temperatures, data summed over the high angle range from $2\theta = 55 - 95^o$ contain only phonon scattering (fig. 2(b)), while the low angle part of the neutron spectrum summed from $1 - 30^o$ contains scattering from both phonons and spin waves (since the magnetic scattering disappears at high angles due to the magnetic form factor). The phonon scattering is removed from the low angle data by subtracting the properly scaled high angle data, as shown in fig 2(c). Figure 2(d) shows that the resulting magnetic intensity for LFO consists of a single peak at ~75 meV. The strong peak at 0 meV is elastic scattering and very weak peaks at ~20 and 30 meV arise from imperfect phonon subtraction. Due to the use of powder samples, the experiment measures the polycrystalline-averaged spin excitation spectrum, with neutron intensity related to the spin wave density-of-states (SWDOS). Figure 2(d) also shows the magnetic scattering from LSFO in the Verwey phase at $T = 10$ K. In LSFO, the SWDOS is split into two bands. The high energy band consists of a single broad peak at ~ 85 meV. The low energy band consists of broad peaks at ~35 meV and ~55 meV. Residual phonon intensity is likely present at ~20 and 30 meV, as well.

Due to predictions that $|J_F/J_{AF}| > 2$ in LSFO [10], neutron spectra were measured up to energy transfers of 250 meV with no additional magnetic scattering observed, as shown in the inset of figure 3. The maximum spin wave energy of ~110 meV in LSFO sets an upper limit on $J_F$, as discussed below. Figure 3 shows the temperature dependence of the spin wave scattering in LSFO. As the temperature is raised, the ~85 meV spin wave band gradually shifts to lower energies. Just below the transition at $T = 200$ K, the spin wave scattering is strongly damped. Spin wave damping may be caused by magnon-magnon interactions near $T_V$, or possibly by charge fluctuations since the optical gap measured by infrared reflectivity closes rapidly near $T_V$ from its maximum value of ~ 130 meV [16]. Above $T_V = 210$ K, spin wave scattering disappears and is replaced by a broad paramagnetic-like scattering. This is also seen in the inset to Fig. 3 with a different experimental configuration.

In LFO, each $Fe^{3+}$ site has a 1/2-filled $3d^5$ shell and nearest-neighbor Fe spins are coupled by strong antiferromagnetic superexchange interactions ($J_{AF} < 0$). Distortions of the perovskite crystal lattice due to rotations of oxygen octahedra reduce the space group symmetry from cubic to orthorhombic (*Pbnm*) giving rise to a very small magnetocrystalline anisotropy [11]. Assumption of isotropic exchange leads to the observed G-type antiferromagnetic structure of LFO (all nearest-neighbor Fe spins antiparallel). In the perovskite structure, next-nearest-neighbor magnetic exchange (and beyond) is significantly weaker than nearest-neighbor exchange and can be ignored. In this situation, it is appropriate to represent the spin dynamics with the Heisenberg model Hamiltonian containing a single exchange parameter,

$$H(LFO) = -J_{AF} \sum_{<i,j>} \mathbf{S}_i^{3+} \cdot \mathbf{S}_j^{3+} \quad (1)$$

where $\mathbf{S}_i$ represents the i*th* iron atom and $<i,j>$ means the sum is only over nearest-neighbors. Using mean-field theory, $J_{AF} \sim -3k_B T_N/zS^{3+}(S^{3+}+1) \sim -3.7$ meV (where $S^{3+} = 5/2$ and $z = 6$). Within linear spin wave theory, we calculate the spin wave dispersion, $\omega(\mathbf{q})$, and SWDOS, $Z(\omega) = \sum_{\mathbf{q}} \delta(\omega - \omega(\mathbf{q}))$. For the G-type LFO magnetic structure, the SWDOS consists of a single sharp peak with an energy of $6|J_{AF}|S^{3+}$ (the zone boundary spin wave energy), leading to the result that $J_{AF} = -4.9$ meV, somewhat larger than the mean field value [11].

To more properly compare Heisenberg model results to the powder neutron data, we obtained spin wave energies and eigenvectors from the model and used them to calculate neutron intensities due to coherent spin wave scattering, S($\mathbf{Q}$,ω) (where $\hbar\mathbf{Q}$ is the momentum transfer) [17]. Polycrystalline-averaging of S($\mathbf{Q}$,ω) was then performed by Monte-Carlo integration over a large number of **Q**-vectors lying on a constant-|**Q**| sphere, $S(|Q|,\omega)$. For LFO, data and calculations (broadened by instrumental resolution) of $S(|Q|,\omega)$ at T = 10 K are compared in figures 4(a) and (b). Figure 4(c) compares the low angle-summed model calculation and data for LFO. Agreement between the model calculation and data are excellent.

The (111) CO pattern in LSFO contains $Fe^{3+}$- $Fe^{3+}$ and $Fe^{5+}$- $Fe^{3+}$ nearest-neighbor pairs, but no $Fe^{5+}$- $Fe^{5+}$ pairs (see fig. 1). Due to the small charge-transfer gap in LSFO, some fraction of doped holes in LSFO reside on oxygen. The exchange between $Fe^{3+}$ and nominal $Fe^{5+}$ ions is ferromagnetic ($J_F$) whether the holes are on iron or oxygen. When the holes are on iron, ferromagnetic superexchange occurs between half-filled and empty $e_g$ orbitals, $Fe^{5+}$ ($3d^3$) - $O^{2-}$ ($2p^6$) - $Fe^{3+}$ ($3d^5$). When a single hole is on oxygen, sharing of the spin-polarized oxygen electron leads to ferromagnetic exchange, $Fe^{4+}$ ($3d^4$) - $O^-$ ($2p^5$) - $Fe^{3+}$ ($3d^5$). The presence of oxygen holes between $Fe^{3+}$ pairs will reduce $J_{AF}$ as compared to the parent insulator LFO. The alternation of ferromagnetic and antiferromagnetic bonds satisfies the observed charge and spin ordering patterns [13,14]. The Heisenberg Hamiltonian for LSFO is then

$$H(LSFO) = -J_{AF} \sum_{<i,j>} \mathbf{S}_i^{3+} \cdot \mathbf{S}_j^{3+} - J_F \sum_{<i,j>} \mathbf{S}_i^{3+} \cdot \mathbf{S}_j^{5+} \qquad (2)$$

where sums are over nearest neighbors of each pair-type. A combination of neutron scattering and Mössbauer measurements estimate the iron valences to be ~$Fe^{3.4+}$ and $Fe^{4.2+}$ due to hybridization with oxygen [14]. Using these results, we assign $S^{3+} \approx 5/2$ and $S^{5+} \approx 2$ in the Heisenberg model for LSFO. The model calculations have best agreement with the data for $J_{AF}$ = -3.5 meV and $J_F$ = 5.1 meV, as shown in fig. 4(d).

By comparison with LFO (Fig. 2(d)), it is clear that ferromagnetic exchange in LSFO splits the single SWDOS peak. The high energy band originates from ferromagnetic-like spin waves centered on the MCDW with a maximum energy of roughly $\sim 3J_F(2S^{3+} + S^{5+}) \sim 110\,\text{meV}$, while the lower band consists of antiferromagnetic spin waves propagating between the domain walls with energy $\sim 3J_{AF}S^{3+} + 3J_F S^{5+} \sim 55\,\text{meV}$. The Heisenberg model calculations do not show quantitative agreement with the LSFO data (fig. 4(d)). In particular, the model does not capture the observed spectral weight near 35 meV, which is consistent with weakened antiferromagnetic bonds in the region between the MCDW. This could occur due to an appreciable oxygen deficiency that would break antiferromagnetic bonds ($<z> < 6$). However, the weight of the 35 meV peak is significant (~15%) and would require a much larger oxygen deficiency than we observe in our samples (~2%). The 35 meV feature could also imply that some antiferromagnetic bonds are weakened due to coupling with charge fluctuations, where the optical electronic gap is ~ 130 meV at low temperatures [16]. The observed closing of the optical gap just below $T_V$ would give rise to stronger coupling, and may explain the temperature dependent softening of the spectrum just below $T_V$ (Fig. 3). In addition, comparison to the model indicates that magnetic features are severely energy broadened by as much as 7 meV beyond the instrumental resolution. Such broadening can indicate strong coupling or a distribution of exchange interactions (due to inhomogeneity, for example).

The reduction of $J_{AF}$ by ~30% in LSFO from its value in the parent compound LFO is consistent with oxygen hole density between $Fe^{3+}$ - $Fe^{3+}$ pairs and inconsistent with scenarios where holes are strongly localized to the domain wall. Based on this, we propose a scenario where the MCDW is a charge density wave, with appreciable hole density in the antiferromagnetic region between the domain walls. Figure 1 illustrates this scenario by showing a schematic drawing of iron and oxygen positions in the [001] plane of LSFO. Oxygens surrounding $Fe^{5+}$ have significant hole density, while oxygens between $Fe^{3+}$ - $Fe^{3+}$ pairs have smaller hole density. Despite having hole density on oxygen, the magnetic domain wall is still centered on the nominal $Fe^{5+}$ metal sites (i.e. it is *not* an oxygen centered domain wall).

The measured exchange ratio $|J_F/J_{AF}| = 1.5$ is greater than one implying that magnetic interactions alone are sufficient for stabilizing the observed (111) structure. It should be noted that for holes are primarily on iron, the (111) structure also has the minimum Coulomb energy. However, the oxygen character of the doped holes and delocalization, as indicated by the presence of doped holes between domain walls, will strongly reduce the influence of the Coulomb energy. Elastic interactions may affect the stability of the CO to a some extent. Although crystalline distortions due to the Verwey transition are very small [13], significant changes of the phonon spectrum related to charge ordering have been observed [16,18]. Thus, it appears plausible that the Verwey transition in LSFO occurs without a dominant influence from Coulomb interactions.

**Acknowledgments** RJM thanks J. Schmalian for useful comments and W. Beyermann for sample characterization. Ames Laboratory is supported by the U. S. Department of Energy Office of Science under Contract No. W-7405-ENG-82. The work has benefited from the use of the Los Alamos Neutron Science Center at Los Alamos National Laboratory. LANSCE is funded by the U.S. Department of Energy under Contract No.W-7405-ENG-36.10

**Figure Captions**

FIG. 1. Schematic diagram of oxygen hole density and iron spins in the (001) plane of LSFO. Open circles denote oxygen and circle radius represents hole density. Black (gray) circles are nominal $Fe^{5+}$ ($Fe^{3+}$) ions. The dotted line indicates a metal-centered domain wall.

FIG. 2. **a**. Inelastic neutron scattering intensity of $LaFeO_3$ (color scale) versus scattering angle and energy transfer at $T = 10$ K and $E_i = 160$ meV. Horizontal white lines delineate regions where phonon and magnetic scattering are isolated. **b**. Neutron intensity summed over the angle range from 55 - 95° originating from phonons. **c**. Neutron intensity summed over the low angle range from 1 - 30° (dots) and phonon background from scaled from high angle sum (magenta hatched region) **d.** Isolated magnetic scattering from LFO (green) and LSFO (red) at $T = 10$ K.

FIG. 3. **a.** Temperature dependence of the magnetic scattering from LSFO with $E_i = 120$ meV. Successive curves are offset by 12 units. (Inset) Magnetic scattering from LSFO up to high energies with $E_i = 300$ meV at $T = 10$ K (blue) and $T = 250$ K (red).

FIG. 4. **a.** Inelastic neutron scattering intensity, $S(|Q|,\omega)$, for LFO at $T = 10$ K and $E_i = 160$ meV. White lines denote limits of constant angle summation, $2\theta = 1 – 30°$. **b.** Calculation of $S(|Q|,\omega)$ for LFO at $T = 10$ K and $E_i = 160$ meV using Heisenberg model with $J_{AF} = -4.9$ meV and $S^{3+} = 5/2$. **c.** Comparison of LFO magnetic scattering data at $T$

= 10 K and Heisenberg model calculation summed from $2\theta = 1 - 30°$. **d**. Comparison of LSFO magnetic scattering data at $T = 10$ K and Heisenberg model calculation with $J_{AF}$ = -3.5 meV, $J_F$ = 5.1 meV, $S^{3+}$ = 5/2, and $S^{5+}$ = 2 summed from $2\theta = 1 - 30°$.

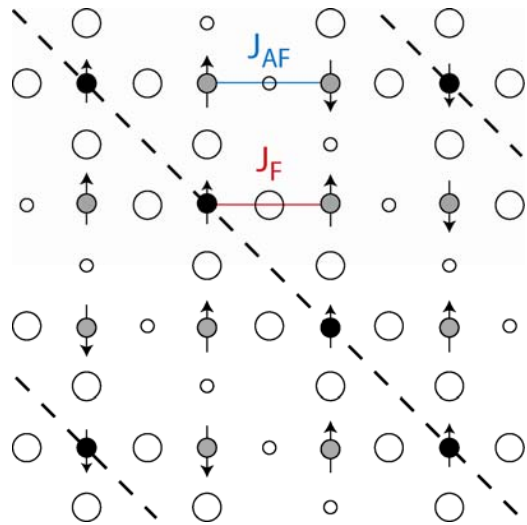

**McQueeney - FIG. 1**

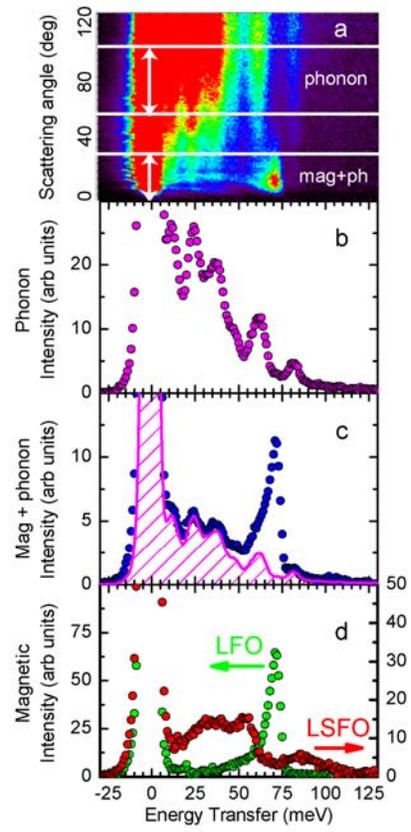

**McQueeney - FIG. 2**

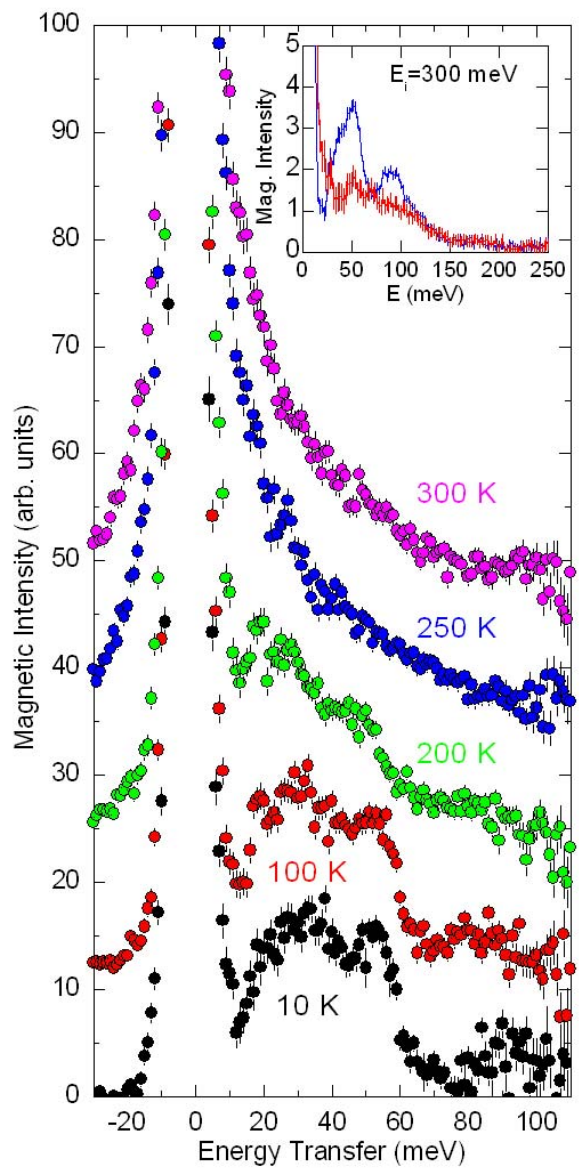

**McQueeney - FIG. 3**

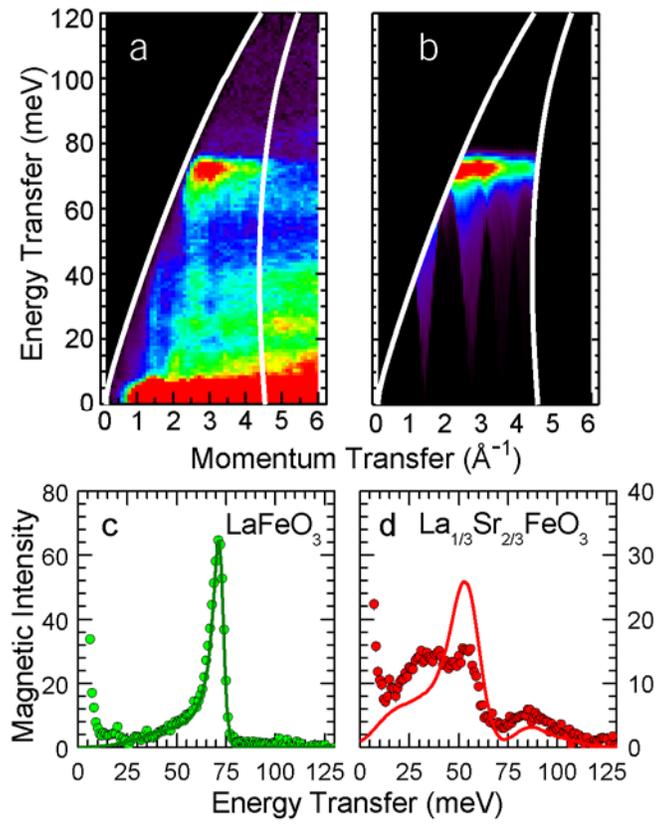

**McQueeney - FIG. 4**